\documentclass[11pt]{article}

\usepackage[parfill]{parskip}  
\usepackage[english]{babel}
\usepackage{amsmath}
\usepackage{amsfonts}
\usepackage{amssymb}
\usepackage{amsthm}
\usepackage[pdftex]{graphicx}
\usepackage{epstopdf}
\usepackage{enumitem}
\usepackage{geometry}
\usepackage{hyperref}
\usepackage{tikz}
\usepackage[all]{xy}
\usepackage{multirow}
\usepackage{anysize}
\usepackage{color}
\usepackage{pdflscape}
\usepackage{lineno}
\usepackage{cite}
\usepackage{lipsum}
\newenvironment{bottompar}{\par\vspace*{\fill}}{\clearpage}

\newcommand{\beq}{\begin{equation}}
\newcommand{\eeq}{\end{equation}}
\newcommand{\commentout}[1]{}

\def\wl{\par \vspace{\baselineskip}}

\marginsize{3.2cm}{3.2cm}{3.2cm}{3.2cm}

\title{Group size effect on cooperation in one-shot social dilemmas II. Curvilinear effect}

\author{Valerio Capraro$^{1,*}$ \& H\'el\`ene Barcelo$^2$}

\begin{document}


\maketitle

\begin{center}
\emph{Forthcoming in PLoS ONE}
\end{center}

\begin{bottompar}
 $^1$Center for Mathematics and Computer Science (CWI), 1098 XG, Amsterdam,  The Netherlands. 

$^2$Mathematical Science Research Institute (MSRI), CA 94720, Berkeley, California, USA.
\wl
* Corresponding author:

Email: V.Capraro@cwi.nl
\end{bottompar}
\pagebreak

\section*{Abstract}
In a world in which many pressing global issues require large scale cooperation, understanding the group size effect on cooperative behavior is a topic of central importance. Yet, the nature of this effect remains largely unknown, with lab experiments insisting that it is either positive or negative or null, and field experiments suggesting that it is instead curvilinear. Here we shed light on this apparent contradiction by considering a novel class of public goods games inspired to the realistic scenario in which the natural output limits of the public good imply that the benefit of cooperation increases fast for early contributions and then decelerates. We report on a large lab experiment providing evidence that, in this case, group size has a curvilinear effect on cooperation, according to which intermediate-size groups cooperate more than smaller groups and more than larger groups. In doing so, our findings help fill the gap between lab experiments and field experiments and suggest concrete ways to promote large scale cooperation among people. 



\section*{Introduction}

Cooperation has played a fundamental role in the early evolution of our societies\cite{KG,Tomasello14natural} and continues playing a major role still nowadays. From the individual level, where we cooperate with our romantic partner, friends, and co-workers in order to handle our individual problems, up to the global level where countries cooperate with other countries in order to handle global problems, our entire life is based on cooperation.

Given its importance, it is not surprising that cooperation has inspired an enormous amount of research across all biological and social sciences, spanning from theoretical accounts \cite{Tr,Ax-Ha,FF03,nowak2006five,Perc10coevolutionary,press2012iterated,perc2013evolutionary,Ca,hilbe2013evolution,Ra-No,capraro2014translucent} to experimental studies \cite{Andreoni1988why,Fischbacher2001people,milinski2002reputation,Frey2004social,Fischbacher2010social,traulsen2010human,apicella2012social,capraro2014heuristics,capraro2014benevolent,capraro2014good,hauser2014cooperating,gallo2015effects} and numerical simulations\cite{Nowak92evolutionary,boyd2003evolution,santos2005scale,perc2008social,roca2009evolutionary,gardenes2012evolution,jiang2013spreading}.

Since the resolution of many pressing global issues, such as global climate change and depletion of natural resources, requires cooperation among many actors, one of the most relevant questions about cooperation regards the effect of the size of the group on cooperative behavior. Indeed, since the influential work by Olson \cite{olson1965logic}, scholars have recognized that the size of a group can have an effect on cooperative decision-making. However, the nature of this effect remains one of the most mysterious areas in the literature, with some scholars arguing that it is negative \cite{olson1965logic,dawes1977behavior,komorita1982cooperative,baland1999ambiguous,ostrom2005understanding,grujic2012three,vilone2014partner,nosenzo2015cooperation}, others that it is positive \cite{mcguire1974group,isaac1994group,haan2002free,agrawal2006explaining,masel2007bayesian,zhang2011group,szolnoki2011group}, and yet others that it is ambiguous \cite{esteban2001collective,pecorino2008group,oliver1988paradox,chamberlin1974provision} or non-significant \cite{todd1992collective,gautam2007group,rustagi2010conditional}. Interestingly, the majority of field experiments seem to agree on yet another possibility, that is, that group size has a curvilinear effect on cooperative behavior, according to which intermediate-size groups cooperate more than smaller groups and more than larger groups \cite{poteete2004heterogeneity,agrawal2001group,agrawal2000small,yang2013nonlinear,cinner2013looking}.
The emergence of a curvilinear effect of the group size on cooperation in real life situations is also supported by data concerning academic research, which in fact support the hypothesis that research quality of a research group is optimized for medium-sized groups \cite{kenna2011critical,kenna2011critical2,kenna2012managing}.

Here we aim at shedding light on this debate, by providing evidence that a single parameter can be responsible for all the different and apparently contradictory effects that have been reported in the literature. Specifically, we show that the effect of the size of the group on cooperative decision-making depends critically on a parameter taking into account different ways in which the notion of cooperation itself can be defined when there are more than two agents.

Indeed, while in case of only two agents a cooperator can be simply defined as a person willing to pay a cost $c$ to give a greater benefit $b$ to the other person \cite{nowak2006five}, the same definition, when transferred to situations where there are more than two agents, is subject to multiple interpretations. If cooperation, from the point of view of the cooperator, means paying a cost $c$ to create a benefit $b$, what does it mean from the point of view of the \emph{other} player\emph{s}? Does $b$ get earned by each of the other players or does it get shared among all other players, or none of them? In other words, what is the marginal return for cooperation?

Of course, there is no general answer and, in fact, previous studies have considered different possibilities. For instance, in the standard Public Goods game it is assumed that $b$ gets earned by each player (including the cooperator); instead, in the N-person Prisoner's dilemma (as defined in \cite{barcelo2015group}) it is assumed that $b$ gets shared among all players; yet, the Volunteer's dilemma \cite{diekmann1985volunteer} and its variants using critical mass \cite{szolnoki2010impact} rest somehow in between: one or more cooperators are needed to generate a benefit that gets earned by each player, but, after the critical mass is reached, new cooperators do not generate any more benefit; finally, it has been pointed out \cite{marwell1993critical,heckathorn1996dynamics} that a number of realistic situations can be characterized by a marginal return which increases linearly for early contributions and then decelerates, reflecting the natural decrease of marginal returns that occurs when output limits are approached.

In order to take into account this variety of possibilities, we consider a class of \emph{social dilemmas} parametrized by a function $\beta=\beta(\Gamma,N)$ describing the marginal return for cooperation when $\Gamma$ people cooperate in a group of size $N$. More precisely, our \emph{general Public Goods game} is the N-person game in which N people have to simultaneously decide whether to cooperate (C) or defect (D). In presence of a total of $\Gamma$ cooperators, the payoff of a cooperator is defined as $\beta(\Gamma,N)-c$ ($c>0$ represents the cost of cooperation) and the payoff of a defector is defined as $\beta(\Gamma,N)$. In order to have a social dilemma (i.e., a tension between individual benefit and the benefit of the group as a whole) we require that:
\begin{itemize}
\item Full cooperation pays more than full defection, that is, $\beta(N,N) - c > \beta(0,N)$, for all $N$; 
\item Defecting is individually optimal, regardless of the number of cooperators, that is, for all $\Gamma < N$, one has $\beta(\Gamma,N)-c < \beta(\Gamma-1,N)$.
\end{itemize}

The aim of this paper is to provide further evidence that the function $\beta$ might be responsible for the confusion in the literature about group size effect on cooperation. In particular, we focus on the situation, inspired from realistic scenarios, in which the natural output limits of the public good imply that $\beta(\Gamma,N)$ increases fast for small $\Gamma$'s and then stabilizes. 

Indeed, in our previous work \cite{barcelo2015group}, we have shown that the size of the group has a positive effect on cooperation in the standard Public Goods game and has a negative effect on cooperation in the N-person Prisoner's dilemma. A reinterpretation of these results is that, if $\beta(N,N)$ increases linearly with $N$ (standard Public Goods game), then the size of the group has a positive effect on cooperation; and, if $\beta(N,N)$ is constant with $N$ (N-person Prisoner's dilemma), then the size of the group has a negative effect on cooperation. This reinterpretation suggests that, in the more realistic situations in which the benefit for full cooperation increases fast for early contributions and then decelerates once the output limits of the public good are approached, we may observe a curvilinear effect of the group size, according to which intermediate-size groups cooperate more than smaller groups and more than larger groups.

To test this hypothesis, we have conducted a lab experiment using a general public goods game with a piecewise function $\beta$, which increases linearly up to a certain number of cooperators, after which it remains constant. While it is likely that realistic scenarios would be better described by a smoother function, this is a good approximation of all those situations in which the natural output limits of a public good imply that the increase in the marginal return for cooperation tends to zero as the number of contributors grows very large. The upside of choosing a piecewise function $\beta$ is that, in this way, we could present the instructions of the experiment in a very simple way, thus minimizing random noise due to participants not understanding the decision problem at hand (see Method).

Our results support indeed the hypothesis of a curvilinear effect of the size of the group on cooperative decision-making. Taken together with our previous work \cite{barcelo2015group}, our findings thus (i) shed light on the confusion regarding the group size effect on cooperation, by pointing out that different values of a single parameter might give rise to qualitatively different group size effects, including positive, negative, and even curvilinear; and (ii) they help fill the gap between lab experiments and field experiments. Indeed, while lab experiments use either the standard Public Goods game or the N-person Prisoner's dilemma, \emph{real} public goods game are mostly characterized by a marginal return of cooperation that increases fast for early contributions and then approaches a constant function as the number of cooperators grows very large - and our results provide evidence that these three situations give rise to three different group size effects.

\section*{Method}

We have recruited participants through the online labour market Amazon Mechanical Turk (AMT) \cite{paolacci2010running,horton2011online,mason2012conducting}. After entering their TurkID, participants were directed to the following instruction screen.

\emph{Welcome to this HIT.}
 
\emph{This HIT will take about 5 minutes and you will earn 20c for participating.} 
 
\emph{This HIT consists of a decision problem followed by a few demographic questions.}  
 
\emph{You can earn an additional bonus depending on the decisions that you and the participants in your cohort will make.} 

\emph{We will tell you the exact number of participants in your cohort later.} 

\emph{Each one of you will have to decide to join either Group A or Group B.} 
 
\emph{Your bonus depends on the group you decide to join and on the size of the two groups, A and B, as follows:}
\begin{itemize}
\item \emph{If the size of Group A is 0 (that is, everybody chooses to join Group B), then everybody gets 10c}
\item \emph{If the size of Group A is 1, then the person in Group A gets 5c and each person in Group B gets 15c}
\item \emph{If the size of Group A is 2, then each person in Group A gets 10c and each person in Group B gets 20c}
\item \emph{If the size of Group A is 3, then each person in Group A gets 15c and each person in Group B gets 25c}
\item \emph{If the size of Group A is 4, then each person in Group A gets 20c and each person in Group B gets 30c}
\item \emph{And so on, up to 10: If the size of Group A is 10, then each person in Group A gets 50c and each person in Group B gets 60c}
\item \emph{However, if the size of Group A is larger than 10, then, independently of the size of the two groups, each person in group A will still get 50c and each person in group B will still get 60c.}
\end{itemize}

After reading the instructions, participants were randomly assigned to one of 12 conditions, differing only on the size of the cohort ($N=3,5,10,15,20,25,30,40,50,60,80,100$). For instance, the decision screen for the participants in the condition where the size of the cohort is 3 was:

\emph{You are part of a cohort of 3 participants.}

\emph{Which group do you want to join?}

By using appropriate buttons, participants could select either Group A or Group B. 

We opted for not asking any comprehension questions. We made this choice for two reasons. First, with the current design, it is impossible to ask general comprehension questions such as ``what is the strategy that benefits the group as a whole'', since this strategy depends on the strategy played by the other players. Second, we did not want to ask particular questions about the payoff structure since this may anchor the participants' reasoning on the examples presented. Of course, a downside of our choice is that we could not avoid random noise. However, as it will be discussed in the Results section, random noise cannot be responsible for our findings. Instead, our results would have been even cleaner, if we had not had random noise, since the initial increase of cooperation and its subsequent decline would have been more pronounced (see Results section for more details).

After making their decisions, participants were asked to fill a standard demographic questionnaire (in which we asked for their age, gender, and level of education), after which they received the ``survey code'' needed to claim their payment. After collecting all the results, bonuses were computed and paid on top of the participation fee, that was \$0.20. In case the number of participants in a particular condition was not divisible by the size of the cohort (it is virtually impossible, in AMT experiments, to decide the exact number of participants playing a particular condition), in order to compute the bonus of the remaining people we formed an additional cohort where these people where grouped with a random choice of people for which the bonus had been already computed.  Additionally, we anticipate that only 98 subjects participated in the condition with N=100. This does not generate deception in the computation of the bonuses since the payoff structure of the game does not depend on $N$ (as long as $N>10$). As a consequence of these observations, no deception was used in our experiment. 

According to the Dutch legislation, this is a non-WMO study, that is (i) it does not involve medical research and (ii) participants are not asked to follow rules of behavior. See http://www.ccmo.nl/attachments/files/wmo- engelse-vertaling-29-7-2013-afkomstig-van-vws.pdf, Section 1, Article 1b, for an English translation of the Medical Research Act. Thus (see http://www.ccmo.nl/en/non-wmo- research) the only legislations which apply are the Agreement on Medical Treatment Act, from the Dutch Civil Code (Book 7, title 7, section 5), and the Personal Data Protection Act (a link to which can be found in the previous webpage). The current study conforms to both. In particular, anonymity was preserved because AMT ``requesters'' (i.e., the experimenters) have access only to the so-called TurkID of a participant, an anonymous ID that AMT assigns to a subject when he or she registers to AMT. Additionally, as demographic questions we only asked for age, gender, and level of education. 

\section*{Results}

A total of 1.195 \emph{distinct} subjects located in the US participated in our experiment. \emph{Distinct} subjects means that, in case two or more subjects were characterized by either the same TurkID or the same IP address, we kept only the first decision made by the corresponding participant and eliminated the rest. These multiple identities represent usually a minor problem in AMT experiments (only 2\% of the participants in the current dataset). Participants were distributed across conditions as follows: 101 participants played with $N=3$, 99 with $N=5$, 102 with $N=10$, 101 with $N=15$, 98 with $N=20$, 103 with $N=25$, 97 with $N=30$, 99 with $N=40$, 97 with $N=50$, 101 with $N=60$, 99 with $N=80$, 98 with $N=100$.

Fig. 1 summarizes the main result. The rate of cooperation, that is the proportion of people opting for joining Group A, first increases as the size of the group increases from $N=3$ to $N=15$, then it starts decreasing. The figure suggests that the relation between the size of the group and the rate of cooperation is \emph{not} quadratic: while the initial increase of cooperation is relatively fast, the subsequent decrease of cooperation seems extremely slow. This is confirmed by linear regression predicting rate of cooperation as a function of $N$ and $N^2$, which shows that neither the coefficient of $N$ nor that of $N^2$ are significant ($p=0.4692, p=0.2003$, resp.). For this reason we use a more flexible econometric model than the quadratic model, consisting of two linear regressions, one with a positive slope (for small $N$'s) and the other one with a negative slope (for large $N$'s). As a switching point, we use the $N=15$, corresponding to the size of the group which reached maximum cooperation. Doing so, we find that both the initial increase of cooperation and its subsequent decline are highly significant (from $N=3$ to $N=15$: coeff $= 0.0187553$, $p=0.00042$; from $N=15$ to $N=100$: coeff $= -0.00177618$, $p=0.00390$). 

\begin{figure} 
   \centering
   \includegraphics[scale=0.80]{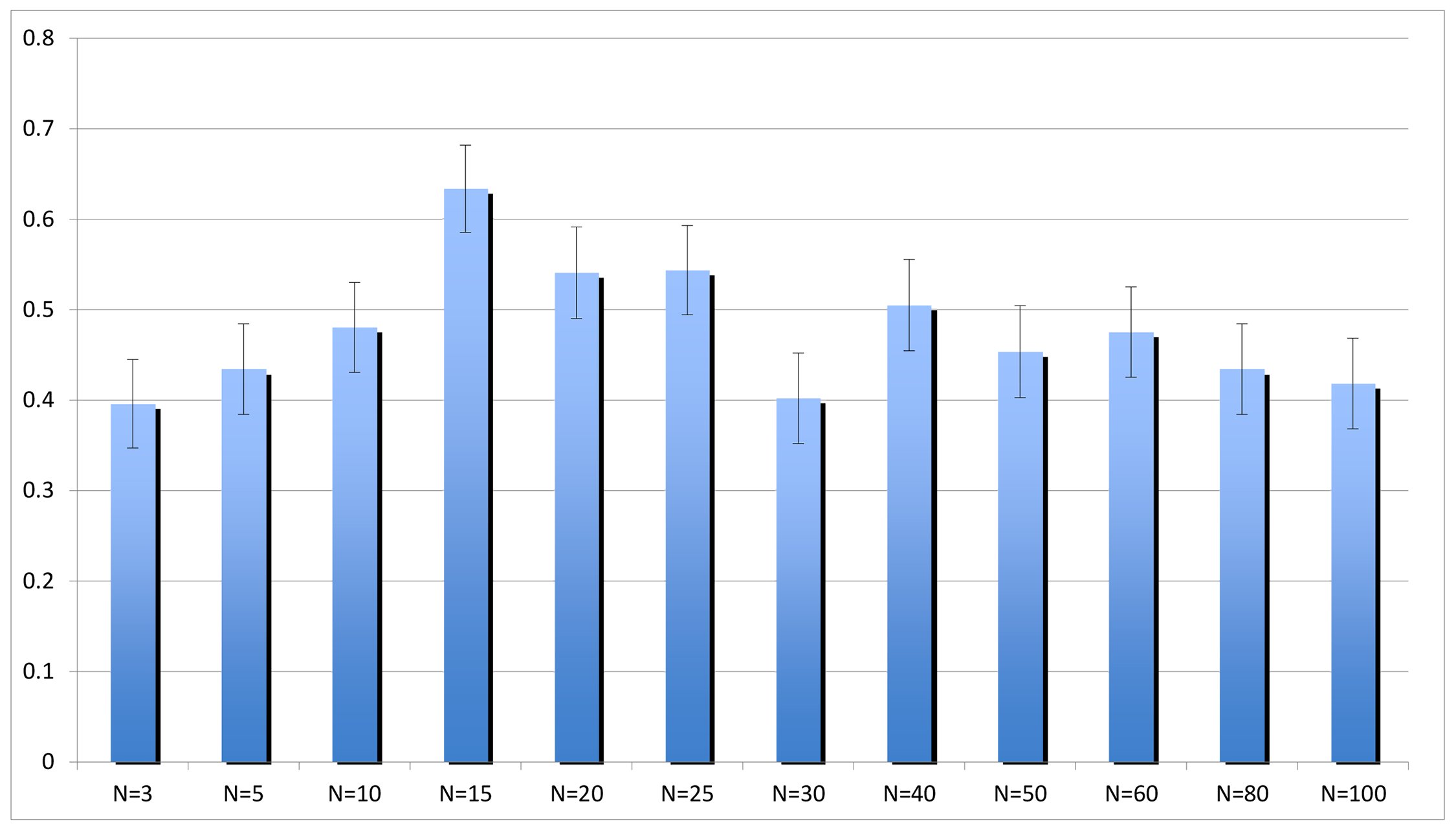} 
   \label{fig:intermediate}
   \caption{\emph{Proportion of cooperators (people choosing to join Group A) for each group size. Error bars represent the standard errors of the means. Group size has initially a positive effect on cooperation, which increases and reaches its maximum in groups of size 15, followed by a gradual decrease. Linear regression predicting cooperation using group size as independent variable confirms that both the initial increase of cooperation and its subsequent decline are highly significant (from $N=3$ to $N=15$: coeff $= 0.0187553$, $p=0.00042$; from $N=15$ to $N=100$: coeff $= -0.00177618$, $p=0.00390$).}}
\end{figure}

We conclude by observing that not only random noise cannot explain our results, but, without random noise, the effect would have been even stronger. Indeed, first we observe that there is no a priori worry that random noise would interact with any condition and so we can assume that it is randomly distributed across conditions. Then we observe that subtracting a binary distribution with average $0.5$ from a binary distribution with average $\mu>0.5$, one would obtain a distribution with average $\mu_0>\mu$. Similarly, subtracting a binary distribution with average $0.5$ from a binary distribution with average $\mu<0.5$ one would obtain a distribution with average $\mu_0<\mu$. Thus, if the $\mu$'s are the averages that we have found (containing random noise) and the $\mu_0$'s are the \emph{true} averages (without random noise), the previous inequalities allow us to conclude that the initial increase of cooperation and its following decrease would have been stronger in absence of random noise.

\section*{Discussion}

Here we have reported on a lab experiment providing evidence that the size of a group can have a curvilinear effect on cooperation in one-shot social dilemmas, with intermediate-size groups cooperating more than smaller groups and more than larger groups. Joining the current results with those of a previously published study of us \cite{barcelo2015group}, we can conclude that group size can have qualitatively different effects on cooperation, ranging from positive, to negative and curvilinear, depending on the particular decision problem at hand. Interestingly, our findings suggest that different group size effects might be ultimately due to different values of a single parameter, the number $\beta(N,N)$, describing the benefit for full cooperation. If $\beta(N,N)$ is constant in $N$, then group size has a negative effect on cooperation; if $\beta(N,N)$ increases linearly with $N$, then group size has a positive effect on cooperation; in the \emph{middle}, all sorts of things may a priori happen. In particular, in the realistic situation in which $\beta(N,N)$ is a piecewise function that increases linearly with $N$ up to a certain $N_0$ and then remains constant, then group size has a curvilinear effect, according to which intermediate-size groups cooperate more than smaller groups and more than larger groups. See Table 1.

\begin{center}
\begin{table}
\begin{tabular}{| l | c | c| }
  \hline                       
  shape of $\beta(N,N)$ & group size effect on cooperation & paper \\
\hline
  linear & positive & Barcelo \& Capraro (2015) \\
  constant & negative & Barcelo \& Capraro (2015) \\
  linear-then-constant & curvilinear & this paper\\
  \hline  
\end{tabular}
\caption{Summary of the different group size effects on cooperation depending on how the benefit for full cooperation varies as a function of the group size.}
\end{table}
\end{center}
To the best of our knowledge, ours is the first study reporting a curvilinear effect of the group size on cooperation in an experiment conducted in the ideal setting of a lab, in which confounding factors are minimized. Previous studies reporting a qualitatively similar effect \cite{poteete2004heterogeneity,agrawal2001group,agrawal2000small,yang2013nonlinear} used field experiments, in which it is difficult to isolate the effect of the group size from possibly confounding effects. In our case, the only possibly confounding factor is random noise due to a proportion of people that may have not understood the rules of the decision problem. As we have shown, our results cannot be driven by random noise and, in fact, the curvilinear effect would have been even stronger, without random noise. Moreover, since our experimental design was inspired by a tentative to mimic all those \emph{real} public goods games in which the natural output limits of the public good imply that the increase of the marginal return for cooperation, when the number of cooperators diverges, tends to zero, our results might explain the apparent contradiction that field experiments tend to converge on the fact that the effect of the group size is curvilinear, while lab experiments tend to converge on either of the two linear effects.

Our contribution is also conceptual, since we have provided evidence that a single parameter might be responsible for different group size effects: the parameter $\beta(N,N)$, describing the way the benefit for full cooperation varies as a function of the size of the group. Of course, we do not pretend to say that this is the only ultimate explanation of why different group size effects have been reported in experimental studies. In particular, in real-life situations, which are typically repeated and in which communication among players is allowed, other factors, such as within-group enforcement, may favor the emergence of a curvilinear effect of the group size on cooperation, as highlighted in \cite{yang2013nonlinear}. If anything, our results provide evidence that the curvilinear effect on cooperation goes beyond contingent factors and can be found also in the ideal setting of a lab experiment using one-shot anonymous games. We believe that this is a relevant contribution in light of possible applications of our work. Indeed, the difference between $\beta(N,N)$ and the total cost of full cooperation $cN$ can be interpreted has the incentive that an institution needs to pay to the contributors in order to make them cooperate. Since institutions are interested in minimizing their costs and, at the same time, maximizing the number of cooperators, it is crucial to understand what is the ``lowest'' $\beta$ such that the resulting effect of the group size on cooperation is positive. This seems to be an non-trivial question. For instance, does $\beta(\Gamma,N)=\frac{\Gamma}{N}\log_2(N+1)$ give rise to a positive effect or is it still curvilinear or even negative? The technical difficulty here is that it is hard to design an experiment to test people's behavior in these situations, since one cannot expect that an average person would understand the rules of the game when presented using a logarithmic functions. 

In terms of economic models, our results are consistent with utilitarian models such as the Charness \& Rabin model \cite{charness2002understanding} and the novel cooperative equilibrium model \cite{Ca,capraro2013cooperative,barcelo2015group}. Both these models indeed predict that, in our experiment, cooperation initially (i.e., for $N\leq10$) increases with $N$ (see \cite{barcelo2015group} for the details), and then starts decreasing. This behavioral transition follows from the simple observation that free riding when there are more than 10 cooperators costs zero to each of the other players and benefits the free-rider. Thus, cooperation in larger groups is not supported by utilitarian models, which then predict a decrease in cooperative behavior whose speed depends on the particular parameters of the model, such as the extent to which people care about the group payoff versus their individual payoff, and people's beliefs about the behavior of the other players. Thus our results add to the growing body of literature showing that utilitarian models are qualitatively good descriptors of cooperative behavior in social dilemmas.

However, we note that while theoretical models predict that the rate of cooperation should start decreasing at $N=10$, our results show that the rate of cooperation for $N=15$ is marginally significantly higher than the rate of cooperation for $N=10$ (Rank sum, $p=0.0588$). Although ours is a between-subjects experiment, this finding seems to hint at the fact that there is a proportion of subjects who would defect for $N=10$ and cooperate for $N=15$. This is not easy to explain: why should a subject cooperate with $N=15$ and defect with $N=10$? One possibility is that there is a proportion of ``inverse conditional cooperators'', who cooperate only if a small percentage of people cooperate: if these subjects believe that the rate of cooperation decreases quickly after $N=10$, they would be more motivated to cooperate for $N=15$ than for $N=10$. Another possibility, of course, is that this discrepancy is just a false positive. In any case, unfortunately our experiment is not powerful enough to detect the reason of this discrepancy between theoretical predictions and experimental results and thus we leave this interesting question for future research.

\section*{Acknowledgements}

V.C. is supported by the Dutch Research Organization (NWO) Grant No. 612.001.352. This material is based upon work supported by the National Science Foundation under Grant No. 0932078000 while the first author was in residence at the Mathematical Science Research Institute in Berkeley, California, during the Spring 2015 semester.

\commentout{
\section*{Figure legends}

Figure 1. Proportion of cooperators (people choosing to join Group A) for each group size. Error bars represent the standard errors of the means. Group size has initially a positive effect on cooperation, which increases and reaches its maximum in groups of size 15, followed by a gradual decrease. Linear regression predicting cooperation using group size as independent variable confirms that both the initial increase of cooperation and its subsequent decline are highly significant (from $N=3$ to $N=15$: coeff $= 0.0187553$, $p=0.00042$; from $N=15$ to $N=100$: coeff $= -0.00177618$, $p=0.00390$).

\pagebreak
\begin{center}
\begin{huge}
\textbf{Supplementary Information}
\wl
\end{huge}
\end{center}

}

\end{document}